\documentclass[manuscript]{aastex631}

\usepackage{float}
\makeatletter
\let\newfloat\newfloat@ltx
\makeatother

\usepackage{amsmath,amssymb,amsfonts}
\usepackage{algpseudocode}
\usepackage{algorithm}

\newcommand{\degree}{^\circ{}}

\received{May 29, 2024}

\submitjournal{PSJ}

\shorttitle{Surface roughness of Didymos and Dimorphos}
\shortauthors{Vincent et al.}

\graphicspath{{./}{figures/}}

\begin{document}

\title{Macro-scale roughness reveals the complex history of asteroids Didymos and Dimorphos.}

\correspondingauthor{Jean-Baptiste Vincent}
\email{jean-baptiste.vincent@dlr.de}

\author{Jean-Baptiste Vincent}
\affiliation{DLR Institute of Planetary Research, Berlin, Germany}

\author{Erik Asphaug}
\affiliation{Lunar and Planetary Laboratory, University of Arizona, Tucson, AZ 85721, USA}

\author{Olivier Barnouin}
\affiliation{Johns Hopkins University Applied Physics Laboratory, Laurel, MD, USA}

\author{Joel Beccarelli}
\affiliation{INAF-OAPD Astronomical Observatory of Padova, Italy}

\author{Paula G. Benavidez}
\affiliation{IUFACyT - DFISTS, Universidad de Alicante, Spain}

\author{Adriano Campo-Bagatin}
\affiliation{IUFACyT - DFISTS, Universidad de Alicante, Spain}

\author{Nancy L. Chabot}
\affiliation{Johns Hopkins University Applied Physics Laboratory, Laurel, MD, USA}

\author{Carolyn M. Ernst}
\affiliation{Johns Hopkins University Applied Physics Laboratory, Laurel, MD, USA}

\author{Pedro H. Hasselmann}
\affiliation{INAF-OAR, Rome, Italy}

\author{Masatoshi Hirabayashi}
\affiliation{Daniel Guggenheim School of Aerospace Engineering, Georgia Institute of Technology, USA}

\author{Simone Ieva}
\affiliation{INAF-OAR, Rome, Italy}

\author{Özgür Karatekin}
\affiliation{Royal Observatory of Belgium, Brussels, Belgium}

\author{Tomáš Kašpárek}
\affiliation{Brno University of Technology, Czech Republic}

\author{Tomáš Kohout}
\affiliation{Department of Geosciences and Geography, University of Helsinki, Finland}
\affiliation{Institute of Geology of the Czech Academy of Sciences, Czech Republic}

\author{Zhong-Yi Lin}
\affiliation{Institute of Astronomy, National Central University, No. 300, Zhongda Rd., Zhongli Dist., Taoyuan City 32001, Taiwan}

\author{Alice Lucchetti}
\affiliation{INAF-OAPD Astronomical Observatory of Padova, Italy}

\author{Patrick Michel}
\affiliation{Université Côte d'Azur, Observatoire de la Côte d'Azur, CNRS, Laboratoire Lagrange, Nice}
\affiliation{The University of Tokyo, Department of Systems Innovation, School of Engineering, Tokyo, Japan}

\author{Naomi Murdoch}
\affiliation{Institut Supérieur de l'Aéronautique et de l'Espace (ISAE-SUPAERO), Université de Toulouse, Toulouse, France}

\author{Maurizio Pajola}
\affiliation{INAF-OAPD Astronomical Observatory of Padova, Italy}

\author{Laura M. Parro}
\affiliation{IUFACyT - DFISTS, Universidad de Alicante, Spain}

\author{Sabina D. Raducan}
\affiliation{University of Bern, Switzerland}

\author{Jessica Sunshine}
\affiliation{Department of Astronomy and Department of Geology, College Park, MD, USA}

\author{Gonzalo Tancredi}
\affiliation{Facultad de Ciencias de la Universidad de la República, Uruguay}

\author{Josep M. Trigo-Rodriguez}
\affiliation{Institute of Space Sciences (CSIC-IEEC), Campus UAB, c/Can Magrans s/n, Cerdanyola del vallès, Barcelona, Catalonia, Spain}

\author{Angelo Zinzi}
\affiliation{ASI-Space Science Data Center, Roma, Italy}

\begin{abstract}

Morphological mapping is a fundamental step in studying the processes that shaped an asteroid surface. Yet, it is challenging and often requires multiple independent assessments by trained experts. Here, we present fast methods to detect and characterize meaningful terrains from the topographic roughness: entropy of information, and local mean surface orientation. We apply our techniques to Didymos and Dimorphos, the target asteroids of NASA’s DART mission: first attempt to deflect an asteroid.
Our methods reliably identify morphological units at multiple scales. The comparative study reveals various terrain types, signatures of processes that transformed Didymos and Dimorphos. Didymos shows the most heterogeneity and morphology that indicate recent resurfacing events. Dimorphos is comparatively rougher than Didymos, which may result from the formation process of the binary pair and past interaction between the two bodies. Our methods can be readily applied to other bodies and data sets.

\end{abstract}


\section{Introduction} \label{sec:intro}
NASA's Double Asteroid Redirection Test (DART) is the first planetary defense mission to attempt the deflection of an asteroid orbit. Its target was Dimorphos, a $151 \pm 5 m~$ diameter S-type asteroid, the secondary component of the 65803 Didymos system (primary's diameter: $761 \pm 26~m$) \cite{daly2023}. The impact successfully occurred on 26 September 2022, at 23:14:24 UTC \cite{daly2023} and changed the orbital period of Dimorphos around Didymos by $-33 \pm 1~min$ \cite{thomas2023}. Shortly before the impact, DART acquired and transmitted high-resolution images of both asteroids. The data revealed complex surfaces with several types of terrains and morphological features like craters and boulders similar to those previously observed on other asteroids \cite{daly2023, barnouin2023}.

The first step in understanding the evolutionary processes that shaped the surfaces of Didymos and Dimorphos consists of mapping the terrains and identifying the different units comprising the overall morphology. This approach is particularly useful for analyzing this binary system as these asteroids will be revisited by ESA's Hera mission in 2027 \cite{michel2022}. The morphological mapping and all information retrieved by DART before and after the impact therefore establish a reference for Hera to identify changes that may have been induced by DART in 2022, or on-going, on both asteroids.

Establishing a morphological map is, however, challenging as the data are limited by the resolution of imaging instruments and the viewing and illumination conditions at the time of observation. The number of images is also limited due to the high-speed nature of the encounter, and only a handful of observations are suited for morphological analysis. Identifying morphological units typically requires multiple experts to independently draw boundaries on surface images until a consensus is reached. While this approach has been successful for previous asteroids (e.g., 25143 Itokawa \cite{fujiwara2006}, 21 Lutetia \cite{thomas2012}, 162173 Ryugu \cite{sugita2019}, 101955 Bennu \cite{barnouin2019}, it is very time consuming and may suffer from human biases that are not easy to quantify. To alleviate this issue, automated methods can be considered to serve as initial guidance and obtain the first assessment of the surface morphology. Examples of automated determination of morphological units on comet 67P/Churyumov-Gerasimenko can be found in \cite{vincent2017} (cliffs) and \cite{thomas2018} (smooth units). Both studies are based on the analysis of the tri-dimensional mesh of the comet's shape model.

This current work focuses on the topographic roughness (a.k.a. multi-scale macro-roughness, for features larger than one pixel). This quantity describes the variability of surface topography as a function of spatial scale. It can be expressed in different ways, like the mean slope angle of the surface with respect to its local neighbourhood, or the root-mean-squared height over a reference surface. A review of techniques commonly used with remote sensing data can be found in \cite{shepard2001}. 

It is well known from previous missions that the surface of a single small body can present a large variability in roughness, typically seen as varying surface texture. Prominent examples are the localised fine-grained deposits on 433 Eros \cite{thomas2002} or the Muses Sea smooth terrain on 25143 Itokawa \cite{fujiwara2006}; both areas contrast with their rougher surroundings. More recent studies, using high-resolution laser altimetry, are available for asteroids Eros \cite{susorney2019} and Bennu \cite{daly2020} and were used to identify regions of different evolutionary states (coupled with crater counting), or assess the thickness of the mobile regolith layer. 

As the roughness is inherently a measure of small-scale amplitude variations in the terrain, its estimate depends on the resolution of available observations and the scale at which it is investigated. At centimeter to meter scale, the roughness might reveal grain size-sorting processes that mobilise the regolith across the surface (e.g. \cite{jawin2022}). At larger measuring scales, it can help to identify craters, boulders, lineaments, or find the boundaries between ponded deposits of fine material and areas of coarser regolith. In turn, those identifications can be associated with gravitational, albedo, compositional maps and so on, to build a full understanding of the surface. 

We note that at pixel-scale and below, the surface roughness is usually retrieved with dedicated photometric modelling of the surface brightness per pixel, through advanced models such as \cite{hapke1984}. See \cite{marshall2018} for roughness studies of comet 67P/Churyumov-Gerasimenko at microwave and infrared wavelengths, or \cite{hasselmann2021} for the optical roughness of asteroid Bennu.

Beyond small bodies, roughness and its connection to morphology have often been studied from topographic measurements of larger bodies, e.g. the Moon \cite{kreslavsky2013} or Mercury \cite{susorney2017}, and show a clear correlation with regolith formation and modification as well as bedrock geology. Roughness is also an important factor to be considered because it affects the balance of radiation and the thermal inertia. The amount and size distribution of the regolith affects the roughness. For example, asteroids smaller than 100~km are often covered by coarse regolith grains that decrease the roughness and thermal inertia \cite{hanus2018}. The situation for binary asteroids could be different as consequence of the gravitational interactions among the bodies. 
~\\
In this work, we present innovative approaches that allow for a fast determination of the surface roughness on airless bodies. Our study focuses on the topographic roughness, in order to identify features that span multiple pixels in the data. We show how techniques derived from information theory can be used to quantify the topographic roughness of asteroid surfaces at multiple spatial scales, and automate the mapping of significant units. We apply our methods to a subset of images of asteroids Didymos and Dimorphos, acquired by DART. We provide a first description of the spatial distribution of roughness on each object, and derive information about the surface evolution of the two bodies.

Our techniques are not constrained to the objects described in this paper and can be readily applied to any other application that need to assess the roughness of a terrain. We are already implementing these techniques in the data analysis pipelines for ESA's Hera mission, that will revisit Didymos and Dimorphos. Our methods may also be considered for autonomous navigation around asteroids or obstacle avoidance during lunar landings.

\section{Methods}\label{sec:method}
We provide here more details on how the roughness measurement techniques work and how they are implemented. Two methods are presented: "Roughness from Texture" (RFT) and "Roughness from Shape" (RFS).

\subsection{RFT: Roughness from texture}\label{sec:method_rft}

The RFT method we developed is inspired by techniques used in Information Theory and defined in \cite{shannon1948}. We measure a parameter known as the \textit{entropy of information} which can be described as the number of bits needed to encode the variability of the signal. Mathematically, it is the log-base-2 of the number of possible outcomes for a message:

\begin{equation}
    H(X) = - \sum_{x \in \chi} p(x) \log_2 p(x)
\end{equation}

For a variable $X$ in the domain $\chi$, we sum over the variable's possible outcomes $x$, and their probability $p(x)$ of occurrence within the signal.

The choice of a logarithm base depends on the application. Here we chose to use $\log_2$, which gives an entropy of information expressed in \textit{bits}. $\log_2$ also has the advantage of being often implemented directly on the hardware (i.e. instruction FYL2X in Intel x86-assembly), and therefore extremely fast to calculate. This is particularly relevant for the use of this technique in embedded software, e.g. for autonomous navigation on board a spacecraft where computing resources are limited.

When applied to imaging data, we measure this parameter in a sliding window of $n \times n$ unordered pixels, across the image. Regions that present a high variability in pixel value will be characterised by a high entropy, while regions with uniform pixel values will have a low entropy.

Photometric models typically consider incident, emission, and phase angles as well as surface albedo and several correcting factors. Typically for asteroids, brightness variation across the surface is mostly a function of the local topography (i.e. incidence and emission). We must however be wary of observations at low phase angles (typically $i, e < 10\degree$) where additional effects such as shadow-hiding and coherent backscattering play a major role and non-topographic parameters like composition and grain size must be accounted for. In a similar fashion, observations at high incidence angle can also be problematic as the long shadows will obscure part of the terrain and also create artificial high-entropy boundaries, where the border between shadowed and illuminated terrain is seen as a large jump in pixel values.
This is, however, a well known problem with the interpretation of imaging data for asteroids, and we preempt it by selecting data and observing conditions that avoid extreme values of $i$ and $e$, as well as masking potentially problematic areas in the images (shadows). With these precautions, our technique relies on the same principles behind digital terrain model reconstruction methods like photoclinometry (a.k.a. "shape from shading"), routinely used by space missions to reconstruct the local terrain.

Therefore, we use the entropy of information (pixel value variability) as a proxy for a measure of the variability of incidence and emission, which is controlled by the local slope of the terrain. This means that the measure of entropy of information informs us directly about the roughness of the surface, at a scale defined by the size of the neighbourhood in which we calculated the entropy.

\begin{figure}[h]%
\centering
\includegraphics[width=0.9\textwidth]{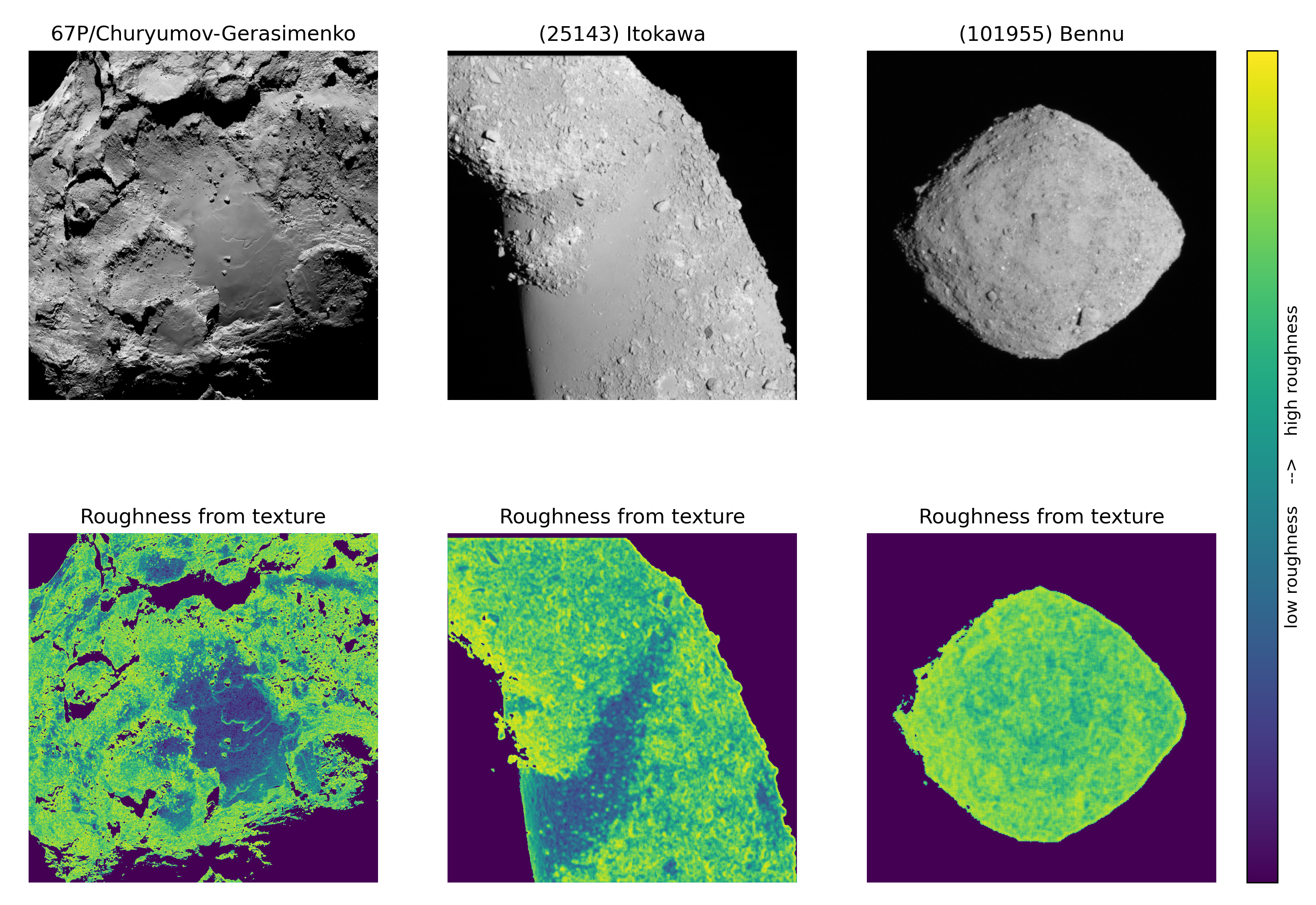}
\caption{Entropy of Information measured on images of other small bodies. The low and high values of entropy match the distribution of rough and smooth terrains on all objects. All images use the same color scale. Qualitatively, 67P and Itokawa display different types of terrains, while Bennu's surface is more uniform. All data is publicly available, see Section \ref{sec:data_av}}
\label{fig:rft_other_bodies}
\end{figure}

Figure \ref{fig:rft_other_bodies} shows examples of entropy measurements on several asteroids and comets, and the clear correlation between the entropy value and the actual roughness of the surface.

\begin{figure}[h]%
\centering
\includegraphics[width=0.9\textwidth]{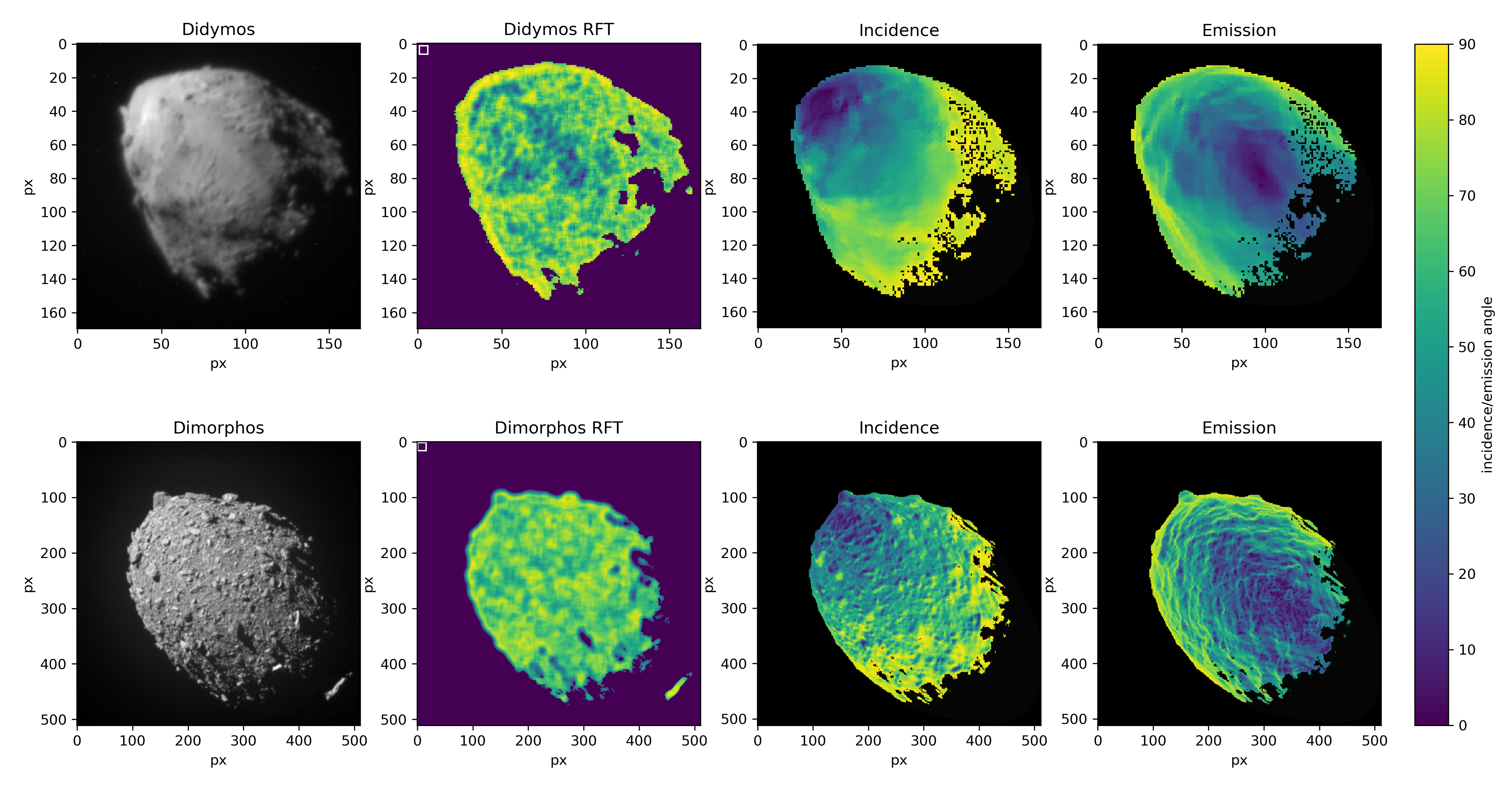}
\caption{Comparison between the Roughness-From-Texture and the illumination angles (incidence and emission) at the time of observation. There is no apparent correlation between these quantities, which suggests that the measurement is robust against a wide range of illuminations.}\label{fig:rft_vs_angles}
\end{figure}

\begin{figure}[h]%
\centering
\includegraphics[width=0.9\textwidth]{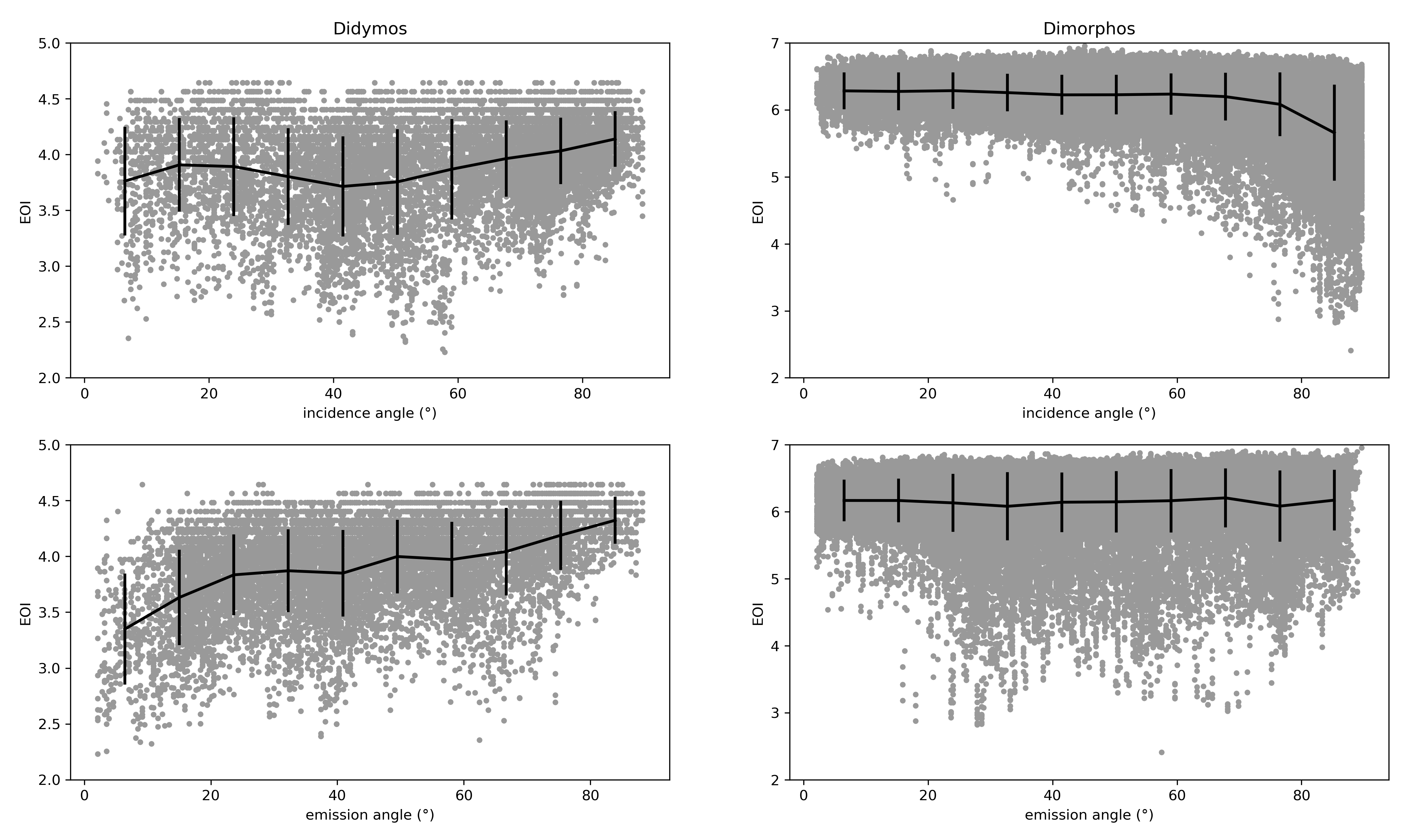}
\caption{Entropy of information vs incidence and emission angles for both asteroids, mean and standard deviations are calculated in bins of $10\degree$. The EOI does not correlate with incidence and emission, except for extreme angles ($i, e  < 20\degree$ and $i, e > 80\degree$) where photometric effects and long shadows start to become significant. }\label{fig:rft_vs_angles2}
\end{figure}

A strong advantage of this method is that it does not require prior knowledge of the photometric response of the surface. More specifically, the results are independent of the viewing angles (incidence, emission, phase) as long as one avoids extreme situations: e.g. at phases below 20°, the brightness variations must account for albedo and material properties. In most images, however, the phase can be ignored as the entropy measure depends on the number of state transitions (how often do pixel values change) rather than the states themselves (absolute difference between pixel values). This means that low-contrast and high-contrast images of the same area have the same entropy, as long as the topography can be detected (again, ignoring opposition and shadows). Figures \ref{fig:rft_vs_angles} and \ref{fig:rft_vs_angles2} shows how incidence and emission angles can essentially be ignored when measuring the roughness with our method on Didymos and Dimorphos, as long as we avoid the more extreme angles.

Besides measuring roughness, our technique can also be used to identify surface changes. By measuring the entropy of the same region at two different epochs, one can determine precisely which areas have seen their texture modified. The technique has been applied successfully to track changes on comet 67P (boulder fields covered/exposed by dust deposition/removal, dunes displacement, formation of circular depressions, etc.), see \cite{vincent2021}.

 We note that the technique is easy to implement and readily available in most image-processing software/libraries. This paper, for instance, used the function available in the open-source Python library \textit{scikit-image} (\url{https://scikit-image.org}). The code implementing the entropy function for this library can be found at \url{https://github.com/scikit-image/scikit-image/blob/main/skimage/filters/rank/generic.py}.

\subsection{RFS: Roughness from shape}\label{sec:method_rfs}

The RFS method operates on a totally different set of data. Rather than considering pixel values in images, we measure the variability of the topography on a reconstructed 3-dimensional model of the object of interest. A shape model is a collection of points in 3D space, linked together in a closed mesh of polygons (typically triangles). For each facet, one can measure its orientation in the body-fixed frame, and compare it to other facets in a given neighbourhood. There are several ways to achieve this, which can be found in the literature and have been applied to Solar System bodies, or everyday objects (e.g. \cite{gowan2023}). The most commonly used technique consists in measuring the RMS distance between facet centers and a reference baseline surface \cite{susorney2019}. Other authors \cite{egorov2017} have also considered comparing the  orientation of a facet in a high-resolution model with the same location on a lower-resolution shape within the same reference frame.

In this paper, we use a slightly different approach which aims to minimize the number of free parameters. Like \cite{egorov2017}, we are interested in the local variation of facet orientation. However, we use the highest-resolution shape model of our target, and do not introduce low-resolution models, to avoid uncertainties that arise from the decimation process. Instead, for each facet, we measure the angle between its normal vector and those of its neighbouring facets. The mean value of this angle is our roughness parameter. To control the scale of the measurement, one can adjust the number of neighbours to be considered. Those can be limited to the immediate three neighbouring triangles, or include all facets within a certain distance of the point of interest. "Distance" can be measured in different ways: shortest path on the surface, latitude/longitude difference, number of neighbours from the starting point, etc.

This technique has the advantage of not requiring any data other than the original shape model. It is purely geometrical, making no assumptions about the internal state of the target (gravity is irrelevant). Because our measurement only depends on the local facet orientation within a given neighbourhood, the results can be directly compared to what is obtained with the RFT method, which also depends on the same parameters.

Of course, we must mention the caveat that our technique is applied to a secondary product (shape model). Hence, the interpretation of our results must account for any uncertainty pertaining to the method used to create that specific model.

The computation can be quite demanding as the file formats used for 3D models of asteroids (e.g. OBJ) do not store information about facet neighbours. Determining if two facets are neighbours requires checking whether they share an edge, i.e. they have two common vertices. A naive way to do this is to load all facets and, for each of them, loop once through the whole list and identify neighbours. One can speed this up by stopping the secondary loop as soon as three neighbours are found, and double-allocating neighbours (if B is a neighbour of A, then A is a neighbour of B). Still, this approach is very time-consuming for models that have millions of facets.

For this work, we have implemented a faster approach to find all neighbours as we load the shape, requiring only one pass through all facets. The algorithm is as follows:

\begin{algorithm}
\caption{Find all neighbours}\label{alg:nb}
\begin{algorithmic}[1]
\Require list of $N$ facets, defined by their vertices: $facet[f] = [i, j, k]$ where $f$ is a facet index and $i, j, k$ are indices of elements in list of vertices 
\State $edge \Leftarrow \textrm{empty array}$
\For{facet index $f$ in range $[0, N]$}
    \State $v_0 \Leftarrow facet[f][0]$
    \State $v_1 \Leftarrow facet[f][1]$
    \State $v_2 \Leftarrow facet[f][2]$
    \State $edge[3 \times f + 0] = [min(v0, v1), max(v0, v1), f]$
    \State $edge[3 \times f + 1] = [min(v1, v2), max(v1, v2), f]$
    \State $edge[3 \times f + 2] = [min(v2, v0), max(v2, v0), f]$
\EndFor
\State Sort $edge$ list by first column
\State For each value of first column, sort $edge$ list by second column
\State $e \Leftarrow 0$
\While{$e < length(edge)$}
    \State $f_a \Leftarrow edge[e][2]$
    \State $f_b \Leftarrow edge[e+1][2]$
    \State Append $f_b$ to $facet[f_a].neighbours$
    \State Append $f_a$ to $facet[f_b].neighbours$
    \State $i \Leftarrow e + 2$
\EndWhile
\end{algorithmic}
\end{algorithm}

The algorithm guarantees that all neighbours are found uniquely, and reading the $edge$ list two rows at a time gives a list of pairs of neighbours.\\

After the neighbours of each facet have been determined, it is trivial to calculate the roughness as a mean orientation angle using simple geometry. For a facet defined by three vertices $[v_0, v_1, v_2]$, the normal vector is given by the cross product between the vectors positioning any pair of clockwise consecutive vertices in the body-fixed frame (e.g. $\overrightarrow{v_0} \times \overrightarrow{v_1}$). The angle $\theta$ between two facets is derived from the dot product between two normal vectors: 
$$
cos(\theta) = 
\frac{\overrightarrow{n_1} . \overrightarrow{n_2}}{||\overrightarrow{n_1}|| \times ||\overrightarrow{n_2}||} $$

~\\
We implemented this approach in the data analysis tool \textit{shapeViewer} \cite{vincent2018}, a scientific software freely available at \url{https://www.comet-toolbox.com}. Our code (\textit{C} language) can process a shape model with 1 million facets in a couple of seconds on a standard laptop (the slowest machine tested has 2GHz CPU, and 4GB RAM). While this is sufficient for our needs, we note that the algorithm could be sped up by parallelizing the construction of the $edge$ list, and selecting a different sorting method (we used  $quicksort$). 

\begin{figure}[h]%
\centering
\includegraphics[width=0.9\textwidth]{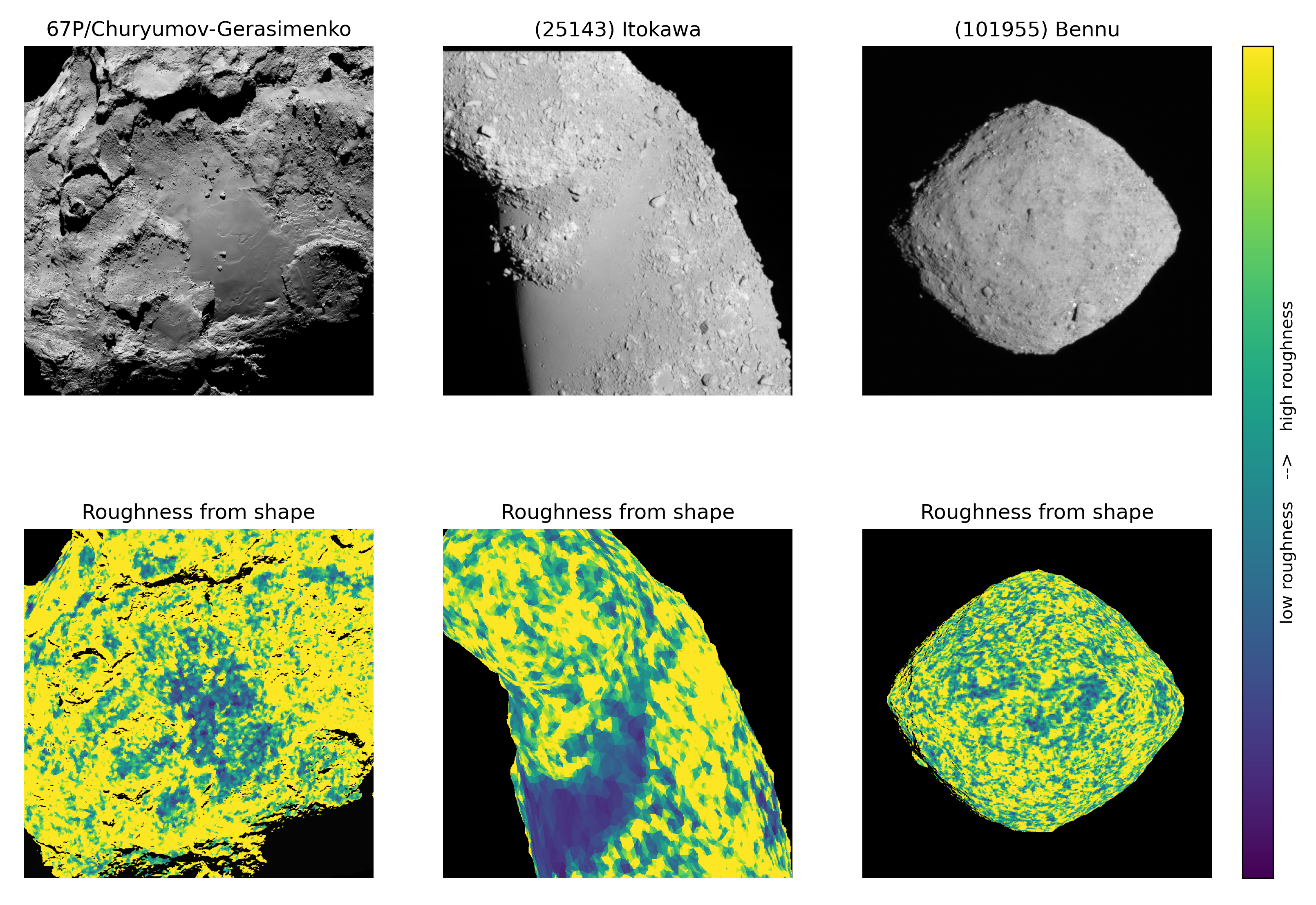}
\caption{Roughness from shape method applied to the reconstructed tri-dimensional models of comet 67P/Churyumov-Gerasimenko and asteroid (25143) Itokawa. Models taken from public archives, processing and rendering done with \textit{shapeViewer} \cite{vincent2018}. We use the same reference images as in Figure \ref{fig:rft_other_bodies}. The smooth/rough terrains on 67P and Itokawa are well identified, within the limitations of the shape model resolution. Bennu does not display much roughness variation.}
\label{fig:rfs_examples}
\end{figure}

Figure \ref{fig:rfs_examples} shows examples of roughness from shape measurements on several asteroids and comets; smooth regions are clearly identified and match those found in Figure \ref{fig:rft_other_bodies} with the RFT method.

\section{Results}\label{sec:result}

\begin{figure}[h]%
\centering
\includegraphics[width=0.9\textwidth]{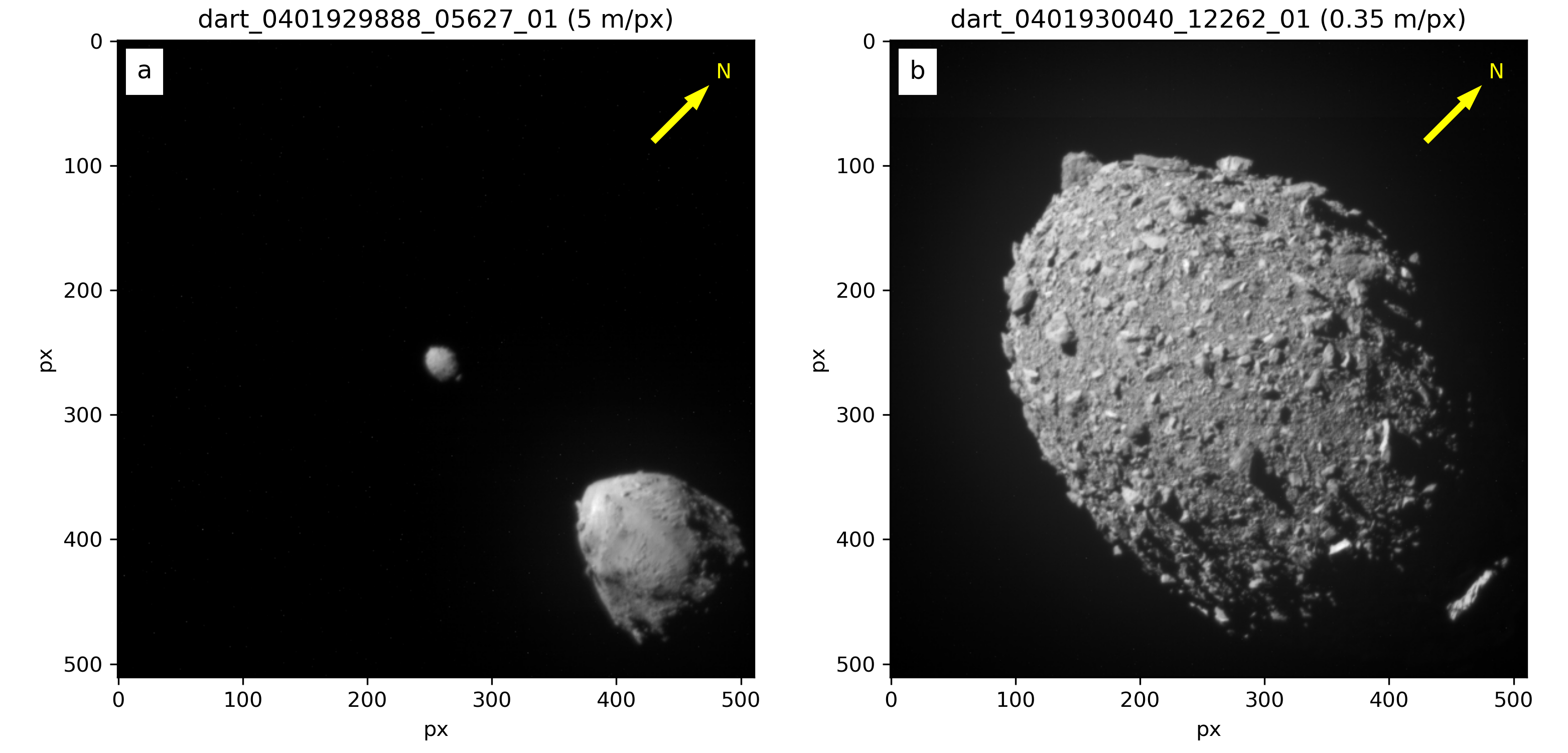}
\caption{DRACO images of Didymos and Dimorphos acquired respectively 2m 43s (a) and 11s (b) before impact. The yellow arrow indicates the direction of the North pole of both asteroids (Didymos's positive spin axis).}\label{fig:reference_images}
\end{figure}

Because the DART spacecraft flew by Didymos and impacted Dimorphos at $6.14~km.s^{-1}$ \cite{daly2023} on an almost straight-on trajectory (only 16.7 degrees from vertical), all images obtained by the Didymos Reconnaissance and Asteroid Camera for OpNav (DRACO, \cite{fletcher2022}) present essentially the same viewing geometry and illumination conditions, with only a change in resolution. In this work, we focus on two specific observations: the highest resolution image of the full sunlit Didymos system (which also contains the highest resolution image of the full sunlit primary), and the highest resolution image of the full sunlit secondary, Dimorphos. Both observations are shown in Figure \ref{fig:reference_images}. We use the calibrated data publicly available in NASA's Planetary Data System archive \cite{ernst2023} with no additional processing.

~\\
We first characterise the surface roughness of both asteroids by measuring the distribution of \textit{entropy of information} (EOI) in these two images. 
In short, the EOI describes the average amount of information contained in a signal \cite{shannon1948}. A rapidly changing signal (either spatially or in time) is considered rich in information (high entropy), while a monotonous signal with little variability is poor in information (low entropy). When applied to imaging data, the signal considered is the distribution of pixel values across the image. The entropy is measured at different scales by considering sub-regions of different sizes in the images, like usually done with spatial filtering \cite{gonzalez2008}. 

This technique has been used for many decades to classify signals or retrieve information in noisy backgrounds, and is readily available in most signal processing or digital image processing software packages. Despite its numerous benefits, we found that the technique has seldom been used in planetary science. To the best of our knowledge, the method has previously only been applied to characterize the roughness of topographic maps derived from laser altimetry (e.g. \cite{li2015}). Our work is the first application of the method to study primary data products such as camera images.

A strong advantage of the method is that it is mostly unaffected by the photometric properties of the surface, as long as we avoid areas with extreme observing and illumination angles. For instance: the opposition effect at low phase angle, or long shadows at high incidence angles. Under typical observing conditions encountered by space missions (i.e. $20^{\circ} <$ phase angle $< 80^{\circ}$) and for uniform albedo surfaces (most asteroids), the entropy is directly a function of the surface topography at the scale of interest. 

The method is presented in detail in \ref{sec:method_rft}, with additional examples for other asteroids and comets. As it is based on the analysis of pixel values in the images, we hereafter refer to the method as roughness-from-texture (RFT), in opposition to alternative methods that are typically based on a shape model.

\begin{figure}[h]%
\centering
\includegraphics[width=\textwidth]{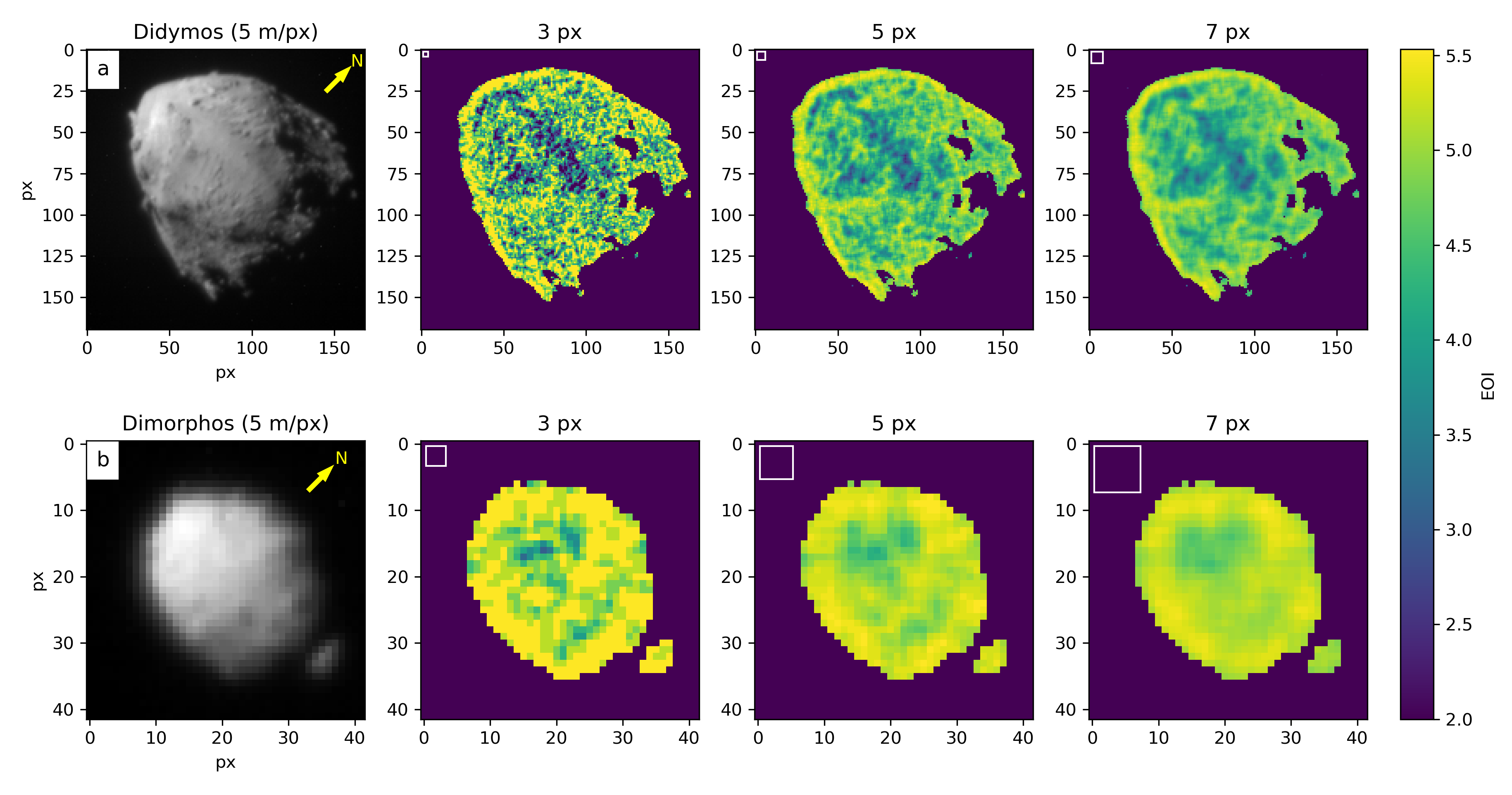}
\caption{Entropy of information (EOI) measured on Didymos [a] and Dimorphos [b] at multiple scales. Image resolution is 5m/px, the measurement scale is given in pixels and also represented visually by a white square in the top-left corner of each panel.}
\label{fig:entropy_5m}
\end{figure}

\subsection{General observations and comparative analysis at 5~m/px}
Figure \ref{fig:entropy_5m} displays a comparative view of the entropy of information on both asteroids at a resolution of 5~m/px. In these images, we measured the EOI at different spatial scales, represented by sliding boxes of $n \times n$ pixels across the images. Here we used scales of 3, 5, and 7 pixels, which correspond to distances of 15, 25, and 35 meters on the surface. All observing conditions (resolution, illumination, viewing angle, etc.) being equal, this allows a direct comparison of the EOI measurement between the two bodies. Of course, at this resolution, Dimorphos is only ~20 pixels wide, while Didymos is five times larger. Still, as long as the entropy measurement scale is smaller than the asteroid, we can get a reliable description of the roughness across the surface.
~\\

There are a couple of artefacts to be aware of when analysing results such as shown in Figures \ref{fig:entropy_5m}, \ref{fig:entropy_d1}, and \ref{fig:entropy_d2}. First, we have masked all shadowed areas and space beyond the asteroid. This is to avoid that low-level noise interpreted as signal. It is possible for very sensitive cameras to actually "see in the dark" when imaging shadowed terrains faintly illuminated by scattered light from nearby lit surfaces. That is however not the case for this dataset. Second, the border between a shadowed and a lit region shows up as a high-entropy edge in our analysis, because pixel values vary by a large amount over a short distance. This leads to an increase of entropy close to the darkest shadows, and around the asteroid limb, and must be taken into account when interpreting the results.

As explained in the introduction, one of the goals of our method is to assist in the definition and characterisation of morphological units. This is achieved by defining thresholds and sorting the surface entropy values in a limited number of bins that correspond to specific morphology. As the entropy values are expressed in \textit{bits} and do not have a direct physical meaning, we first consider the typical entropy values of areas that would be unambiguously characterised as smooth/rough by trained morphology experts, in the data being investigated. We then segment the images according to these thresholds. For bodies whose surfaces shows starker differences in roughness level, this process could be fully automated by identifying peaks in the histogram of entropy values. 

Our results are shown in Figures \ref{fig:entropy_d1} and \ref{fig:entropy_d2}, and discussed hereafter:

\begin{itemize}
\item Both asteroids display significant variations in entropy of information, unrelated to the observing conditions (no correlation between roughness values and incident, emission, or phase angles, see \ref{sec:method_rfs} for additional information). This is indicative of varied roughness levels at the scales considered.
\item Both asteroids show a non-random distribution of smooth and rough terrains. We detect patterns that persist at multiple scales and may indicate boundaries between morphological terrains, or sub-surface structures (see \ref{sec:Didymos}, \ref{sec:Dimorphos}). 
\item Didymos displays a latitudinal trend in roughness values, with the equatorial regions appearing smoother than the mid and high latitudes. Such a latitudinal trend is not detected on Dimorphos.
\item The equatorial regions of Didymos are the smoothest terrains found in the system.
\item Across the scales considered, Dimorphos' surface displays, on average, the same level of roughness as the rougher terrains on Didymos.
\end{itemize}

We now look at each object in more detail, and discuss the interpretation of entropy of information at the highest available resolution. 

\subsection{Didymos}\label{sec:Didymos}
\begin{figure}[h]%
\centering
\includegraphics[width=\textwidth]{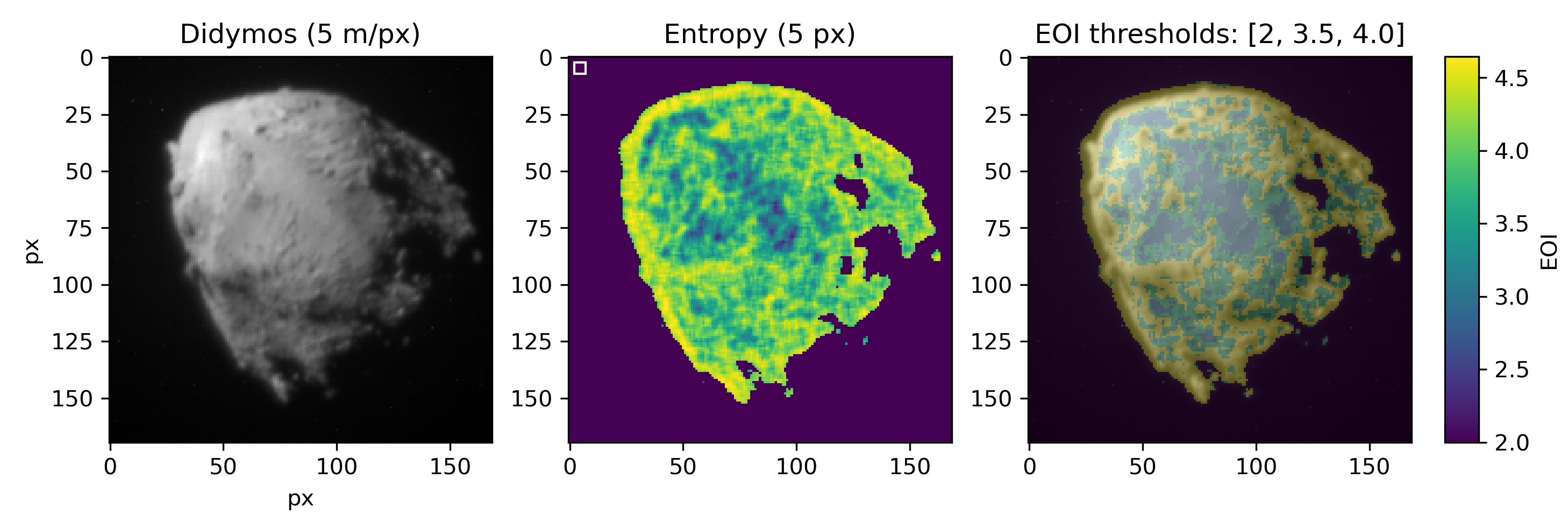}
\caption{Entropy of information (EOI) measured on Didymos at 5~m/px. The thresholds in EOI values have been chosen to enhance the location of smooth, intermediate, and rough terrains on the asteroid.}
\label{fig:entropy_d1}
\end{figure}

\begin{figure}[h]%
\centering
\includegraphics[width=\textwidth]{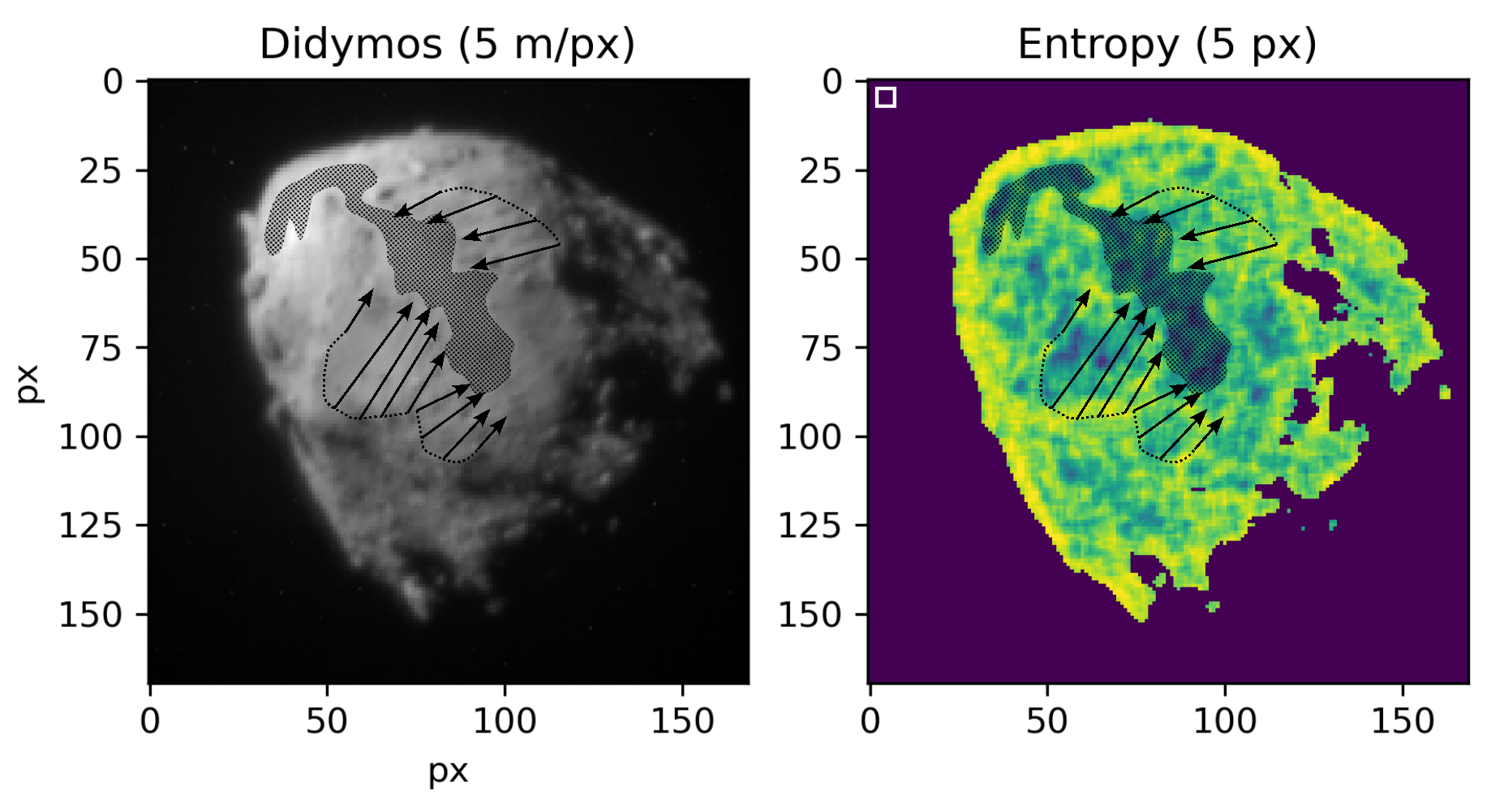}
\caption{Example of geological units detected on Didymos, overlayed on the original image and the map of EOI. The dotted area shows one the contiguous smooth equatorial region. The terrains marked with a dotted line are putative mass wasting break lines, from which material slumped downhill, creating smooth areas bordered at their bottom end by a pile of debris. Whether each mass wasting intercepts another provide information on their time sequence. A complete description of the geological units is available in \cite{barnouin2023}.}
\label{fig:didy_geo}
\end{figure}

Didymos shows unambiguous patterns in its distribution of topographic roughness, with its smoothest regions being located close to the equator (Fig. \ref{fig:entropy_d1}). This is particularly visible in the thresholded entropy image that shows a factor two difference between the entropy level at higher latitudes and the equatorial measurements. We find that the smooth areas do sometimes expand to higher latitudes, particularly in the southern hemisphere, into terrains that resemble avalanche paths. Those lobate features end with a rough "tongue"-like feature that overlaps the equatorial smooth areas. When thresholding the image, we found it useful to introduce three levels of roughness which accurately map the smooth terrains, rough terrains, and intermediate zones that would correspond to the edges of the landslides mentioned previously. Our determination of morphological units matches the geological investigation \cite{barnouin2023}. This is illustrated in Figure \ref{fig:didy_geo} for a couple of morphological units: smooth regions and putative mass wastings.

This mapping is consistent with our understanding of the current state of Didymos. With a rotation period of 2.26 hours \cite{naidu2020}, and an estimated mass of $5.6 \times 10^{11}~kg$ \cite{barnouin2023}, the centrifugal force at the equator of Didymos is comparable in magnitude to the gravitational acceleration ($\simeq 2.2 \times 10^{-4}~ m.s^{-1})$. This means that the surface at low and middle latitudes may be only held together by non-zero mechanical strength and is likely to be easily mobilised, even by the smallest perturbations. Additionally, several authors (e.g., \cite{walsh2012, zhang2021}) have proposed that top-shaped asteroids spinning at a short spin period may structurally fail on their surface if there is a mechanically strong interior. If there is no such strong interior, then the interior fails first, the deformation mode triggering surface failure, causing catastrophic disruption \cite{hirabayashi2022}. However, recent work also shows that the spin period also changes the failure mode \cite{hirabayashi2019}. Surface layers may be more sensitive at a longer spin period, given almost zero cohesion; on the other hand, interiors may be more sensitive at a shorter spin period, needing higher mechanical strength \cite{hirabayashi2020}. In summary, these findings suggest that while Didymos's highly oblate shape likely came from internal failure \cite{barnouin2023}, its current surface can also be sensitive to mass wasting without internal failure, given the existence of internal mechanical strength to support the current shape.

The equatorial regions are, therefore, areas where we expect to find landslides, but also where dust might be able to stick to the surface better than boulders (cohesion of granular matter increases with decreasing particle size \cite{heim1999, scheeres2010, sanchez2014}. The spatial distribution of roughness level, as shown in Fig. \ref{fig:entropy_d1} indicates that multiple regions overlap each other. They display the characteristic features of mass wasting: a sharp upper edge where the slope initially failed, a smooth flow area with few boulders that did fully reach the end of the run, and a rough talus at the bottom of the slope as an accumulation of the mass wasted material. These areas appear to be preferentially oriented from higher latitudes toward the equator, which is consistent with the model predictions. We present an few of the afore mentioned morphological features in Figure \ref{fig:didy_geo}, overlayed on an image of Didymos and on the roughness map. A complete description of the various geological units is available in \cite{barnouin2023}, and a detailed study of boulders rolling down putative avalanche runs can be found in \cite{bigot2023}. Here we emphasise that the roughness analysis provides additional insight which can help refine the geological interpretation.

Mobilized material close to the equator may leave the surface and be put in orbit, depending on the actual asteroid mass and equatorial extents \cite{trogolo2023}. The overlap between these mass-wasted areas indicates that several events took place sequentially, and it is possible that regolith mobilization will occur again on Didymos in the near future, perhaps to be detected by the upcoming Hera mission. The ejecta mass has been measured to be at least $6 \times 10^6~kg$ \cite{moreno2023}, potentially up to $2.2 \times 10^7~kg$ \cite{graykowski2023}. Some of this material may have impacted Didymos, triggering additional mass wasting.

\subsection{Dimorphos}\label{sec:Dimorphos}
\begin{figure}[h]%
\centering
\includegraphics[width=\textwidth]
{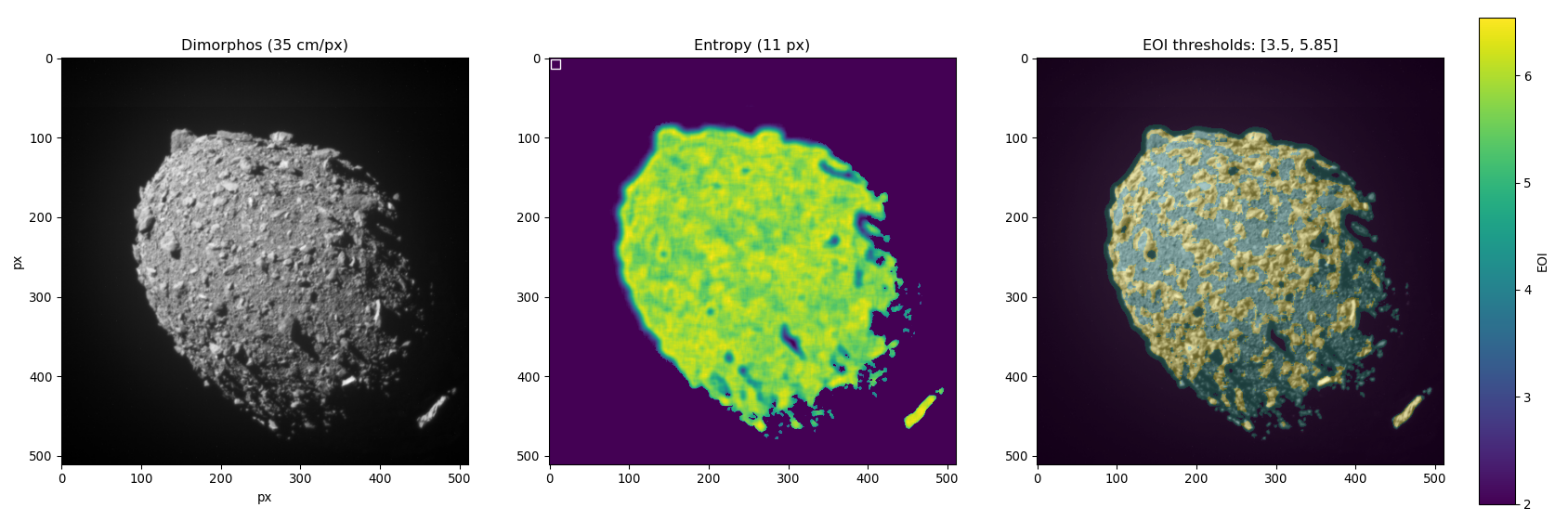}
\caption{Entropy of information (EOI) measured on Dimorphos at 35~cm/px. The thresholds in EOI values have been chosen to clearly separate boulders from pebbles.}
\label{fig:entropy_d2}
\end{figure}

By contrast with its larger companion, Dimorphos presents a much more homogeneous surface, on average rougher than Didymos' at all scales (Figs. \ref{fig:entropy_5m}) and \ref{fig:entropy_d2}). We do not find any smooth region on the surface, which is dominated by high-entropy terrains. However, the EOI is not randomly distributed and sorting the values in two bins leads to a clear separation between boulders and regolith at the chosen scale. Figure \ref{fig:entropy_d2} shows an example of this categorisation, where meter-size boulders are clearly identified by the entropy measurement. This detection is consistent with the distribution reported in \cite{pajola2023}, where the authors manually mapped 4734 boulders on the surface. 

The efficient detection of meter-size objects is due to two supporting factors (1) we selected an entropy window size smaller than the objects, and therefore the filter acts as an edge detector; (2) there is an intrinsic difference in roughness between boulders and regolith substrate, at the resolution considered. This indicates that boulders are rather monolithic entities, rather than agglomerates of pebbles with a same size distribution as the surrounding pebbles.

We stress that for all objects, the entropy of information provides a reliable qualitative estimate of the distribution of smooth and rough terrain, as well as morphological features of interest within those terrains. One must however be cautious about quantitative comparisons, which can only be performed when all measurement parameters are kept constant. We particularly emphasise that it is critical to use objects or areas images at similar resolution, and the entropy is measured at the same spatial scale. In general, we recommend to prioritise the use of high resolution images, even when measuring roughness at large spatial scales. The low resolution may hide topographic features.

At 5~m/px, we do not find any terrain on Dimorphos that would be as smooth as the equatorial regions of Didymos, possibly indicating the absence of small particles, although this cannot be proven from the low-resolution images only. Higher-resolution images of the impact site obtained in the last seconds confirm that the surface of Dimorphos does not have  expansive smooth deposits where there were regolith particles that were unresolved at 5.5~cm/px \cite{daly2023}. 

We acknowledge that DART observed mostly the leading side of Dimorphos (with respect to its orbit around Didymos). Other regions, in particular the areas facing toward or away from Didymos may display a different morphology or boulder distribution, possibly affected by tides induced by the primary. These terrains will be observed by the European Space Agency's Hera mission, scheduled to launch in October 2024.

\subsection{Roughness from shape}
As we measure the entropy on the calibrated images, with no further processing, one may have to consider the potential distortions induced by the asteroid shape itself: a pixel of an area with high emission angle covers, for instance, a much larger surface than a pixel at nadir. This effect must be taken into account when discussing roughness per area, and can be corrected using ortho-rectified images (i.e. images projected onto a common reference frame). This is particularly important if one wants to study the evolution of roughness over time, with multiple images of the same region that may have been acquired from different viewpoints \cite{vincent2021}. Here, we are less concerned about the effect of the shape, as we work on single images and do not compare several viewing observations. We also show in Sec. \ref{sec:method_rfs}, Fig. \ref{fig:rft_vs_angles} that, for this data set, the viewing and illumination angles of singular pixels do not affect the distribution of entropy values.

Still, in order to validate our approach, we also measured the roughness using a more traditional method of directly comparing the topography with a reference baseline, as done in other studies. However, see section \ref{sec:method_rfs}, we do approach this measurement in a new way, in which the baseline is not constrained by free parameters, but is rather directly calculated on the shape model itself. Our roughness from shape (RFS) measures the mean angle between the orientation of a facet on the shape model and its neighbours. The size of the neighbourhood determines the spatial scale at which the roughness is measured.

We based our measurements on the highest resolution shape models obtained by the DART team (Didymos: \cite{palmer2022}, Dimorphos: \cite{daly2023}), and publicly available (see \ref{sec:data_av}). The models have a ground sample distance (separation between vertices) of 25 cm for Dimorphos, and 1.2 m for Didymos, and were resampled to a resolution comparable to the smallest scale at which we measured the entropy of information in the images (25 m for Didymos and 3.85 m for Dimorphos), using the open-source software Meshlab (\url{https://www.meshlab.net/}) and its implementation of the resampling algorithm provided in \cite{garland1997}, which preserves the original orientation of the surface.

\begin{figure}[h]%
\centering
\includegraphics[width=0.9\textwidth]{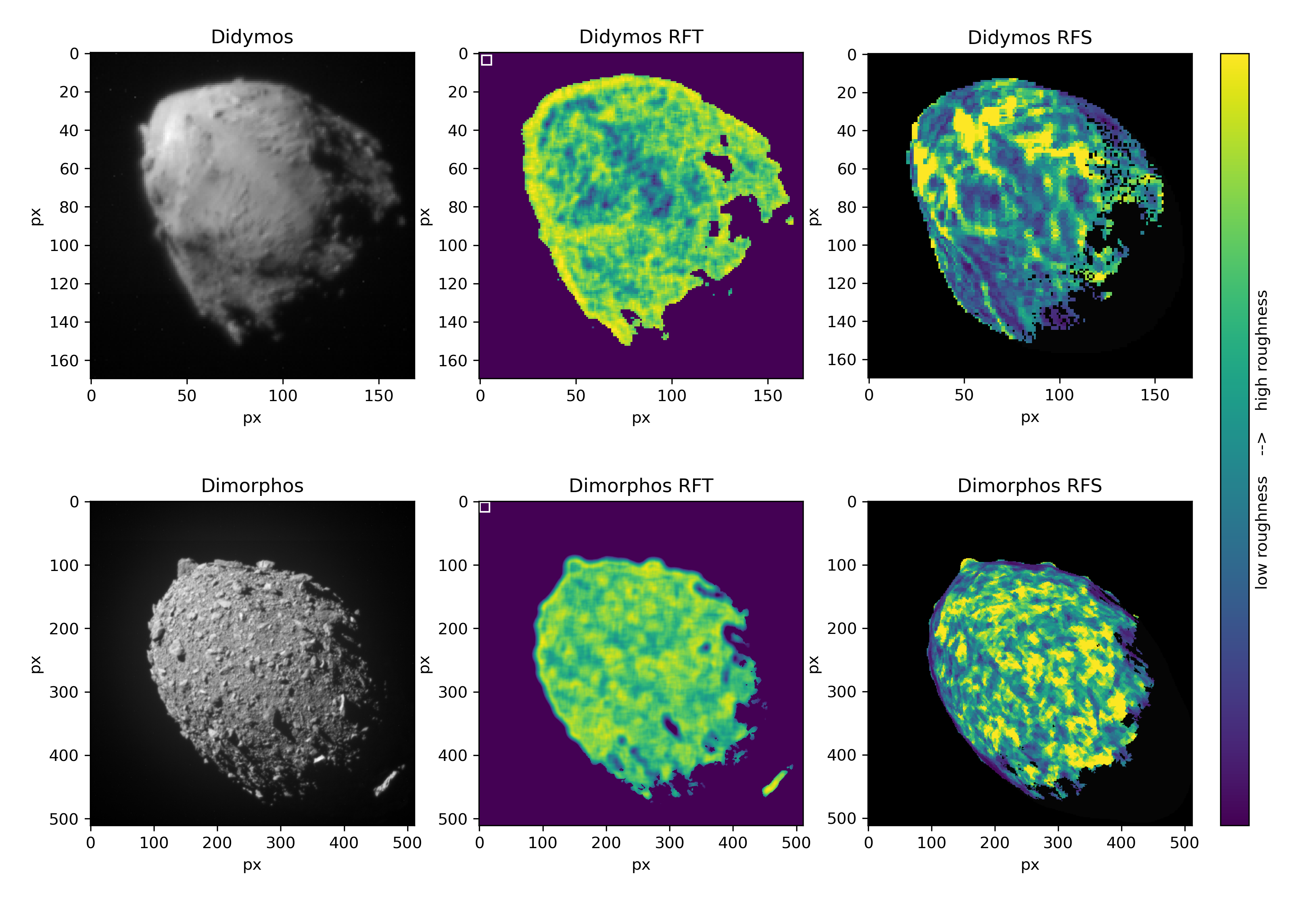}
\caption{Comparison between Roughness-From-Texture and Roughness-From-Shape. Colors indicate qualitatively the distribution of smooth and rough terrains. Both methods lead to a similar description of the surfaces of Didymos and Dimorphos.}\label{fig:rft_vs_rfs}
\end{figure}

We find that the roughness calculated on the shape model (Figure \ref{fig:rft_vs_rfs}), matches very well what we obtained with the RFT. In particular for Dimorphos, the dichotomy between detected in the entropy (boulders vs. regolith) also appears clearly with the RFS method: boulders are detected as areas with rapidly changing local angles on the digital terrain model, which translates into high-entropy regions in the images.

The match is not as good on Didymos. This is mostly due to the fact that the tri-dimensional shape reconstruction of this object is far more challenging, because of the limited data available. Essentially, DART returned only a couple of images of Didymos at a resolution $< 5m/px$, with very little variation in viewing geometry. This limits the scale at which surface features can be reliably reconstructed. On a large scale, though, the analysis of the digital shape model confirms the presence of smoother areas in the equatorial regions, and rougher terrains at higher latitudes.

Overall, we find that both methods converge qualitatively toward a similar description of the asteroid surfaces. In general, for a quantitative assessment, all roughness analysis should be performed on high resolution 3D shape models. Obviously, that is not always possible. Reconstructing accurately the shape of an asteroid from space images usually requires weeks of effort, and large computer resources. Here we have shown that a technique like the EOI provides a reliable first description of the surface and identification of major terrains and landforms of interest like boulders for a much lower cost (fraction of a second). This has implications for on-board autonomous navigation systems, for instance, which do not have the processing resources to work with 3D models.

\section{Discussion}\label{sec:discuss}

In summary, we have analysed images returned by NASA's DART mission via novel approaches that aim to provide a fast identification of different terrain types. We use our method to derive a first characterization of the topographic roughness of asteroids Didymos and Dimorphos.

We show that Didymos and Dimorphos exhibit complex geomorphological features. Didymos' morphology presents large-scale patterns in its roughness distribution, which clearly identify several distinct morphological units: smooth equatorial terrains, rough highlands at high latitudes, and possible landslides flowing from mid-latitudes toward the equator. This pattern is consistent with our current understanding of evolutionary processes that may occur on the surface of rapidly spinning asteroids. The layering of multiple landslides indicates that such events took place sequentially at different epochs and may be related to the spin-up process that led to the current state. This interpretation is compatible with the geological study presented in \cite{barnouin2019}, which concludes that Didymos is structurally weak object with a shear strength in the range 1-10 Pa from surface to interior. This implies that the surface would be extremely prone to failure from the smallest solicitation, and mass wasting is likely to have taken place, and perhaps still occur at the present day.

Dimorphos, on the other hand, presents a much more homogeneous surface, on average rougher than Didymos. We note a clear bi-modal trend in the roughness distribution, which separates decimeter-scale and larger boulders from the finer regolith substrate. The roughness analysis finds no concentrated regions of dust on the surface of Dimorphos, which is confirmed by high-resolution images \cite{daly2023, barnouin2023}. Note that this does not mean that Dimorphos does not host dust at all, but that small particles are not present on the surface in large deposits. Telescopic observations, which are usually sensitive to $µm$-level particles, captured the ejecta plume consisting of such small particles. It is hard to consider that the DART impact generated all such dust particles because highly effective fragmentation only occurs at the impact site, meaning that the large amount of dust observed by telescopic campaigns may have existed before the impact, possibly stored in subsurface regions. 

The lack of dusty regions on the surface of Dimorphos, though, suggests that the binary pair was formed relatively recently, within a time scale shorter than what is needed to create new dust from boulder erosion by meteoroid impact or thermal fatigue (0.1 MYrs, \cite{lucchetti2023}). That is compatible with cratering studies, which indicate that the surface age of Dimorphos is significantly younger than Didymos', in the range 0.09-11 MYrs \cite{barnouin2023}. We also note that a detailed analysis of the morphology of Dimorphos boulders shows that they are primarily fragments of a catastrophic disruption and do not display evidence for significant subsequent erosion \cite{robin2023}. Whether this disruption resulted from YORP spin-up or from collisions cannot be determined by our analysis alone. However, dynamical studies \cite{campo-bagatin2023} strongly suggest that collisions have been the dominant factor driving surface evolution.

As Didymos's equatorial regions appear much smoother than Dimorphos, we suggest that these areas are likely to be covered in material finer than what we detected on Dimorphos at the highest resolution (a few centimetres).

In conclusion, several lines of evidence combine to build a picture of how the Didymos binary system came to be: the rough surface of Dimorphos, the smooth areas lacking boulders on Didymos' equatorial regions, and the relative age difference between the two bodies, may be indicative of a formation process by which Dimorphos is preferentially formed from material that was shed from the primary asteroid.

The methods presented in this paper provide a novel way to assess the surface roughness of an airless body, from texture alone, or from a shape model. In both cases, our approach strive to obtain a rapid evaluation of the surface properties while minimising the number of free parameters. We find that the roughness from texture, based on measuring the entropy of information in images, is particularly robust and low in resources. This make it a good candidate for embedded software that could support autonomous navigation, for instance with future asteroid missions or a lunar lander. We are currently implementing the technique in the data analysis pipeline for ESA's Hera mission, which is scheduled to launch in 2024 and will fully characterise asteroids Didymos and Dimorphos at high resolution in 2026.


\section{acknowledgements}\label{sec:akn}
This work was supported by the DART mission, NASA Contract No. 80MSFC20D0004. It has also received funding from the European Union’s Horizon 2020 research and innovation programme under grant agreement No 870377 (project NEO-MAPP) and the Centre National d’Etudes Spatiales (CNES). AL and MP acknowledge financial support from Agenzia Spaziale Italiana (ASI-INAF contract No. 2019-31-HH.0 and No. 2022-8-HH.0). ACB acknowledges funding by EC H2020-SPACE-718 2018-2020 / H2020-SPACE-2019, and by MICINN (Spain) PGC2021, PID2021-125883NB-C21. JMT-R acknowledges financial support from the project PID2021-128062NB-I00453 funded by MCIN/AEI/10.13039/501100011033. TK is supported by Academy of Finland project 335595 and by institutional support RVO 67985831 of the Institute of Geology of the Czech Academy of Sciences.

\appendix

\section{Data Availability}\label{sec:data_av}
All data used in this article is publicly available on NASA's Planetary Data System archive (\url{https://pds-smallbodies.astro.umd.edu/}), including images, associated documentation and advanced products like the asteroid shape models.
Reference images used in this study: 67P: N20140903T034422640ID20F22, Itokawa: st\_2474846738\_v, Bennu: 20181113T045713S212\_pol\_L1pan, Didymos: dart\_0401929888\_05627\_01\_iof, Dimorphos: dart\_0401930040\_12262\_01\_iof

\section{Code Availability}\label{sec:code_av}
The image analysis performed in this paper is based on publicly available python libraries, all links have provided in the text. The shape model analysis is based on custom C code added to the publicly available tool shapeViewer, \citep{vincent2018}. Because the code relies on other parts of the software to load the shape model and display the results on the screen, we opted for providing the detailed algorithm in section \ref{sec:method_rfs} rather than the actual implementation. The compiled software is freely available to verify our results. Code can be made available on request.

\bibliography{main}{}
\bibliographystyle{aasjournal}


\end{document}